\documentclass[aps,prl,twocolumn,showpacs]{revtex4}
\usepackage{amsmath}
\usepackage{subfigure}
\usepackage{graphicx}
\usepackage{epsfig}
\usepackage{txfonts}
\usepackage{epsf}

\newlength{\upit}\upit=0.1truein

\newcommand{\ltappr}{{{\lower4pt\hbox{$<$} } \atop \widetilde{ \ \ \ }}}
\newlength{\bxwidth}\bxwidth=1.5 truein

\newcommand{\bk}{{\bf k}}
\newlength{\figwidth}
\newlength{\shift}
\usepackage{epsf}

\newcommand \bea {\begin{eqnarray} }
\newcommand \eea {\end{eqnarray}}

\newcommand{\beg}{\begin{equation}}
\newcommand{\en}{\end{equation}}

\newcommand{\dg}{^{\dagger }}

\begin{document}

\title{Topological Kondo Insulators}
\author{Maxim Dzero$^{1}$, Kai Sun$^1$, Victor Galitski$^1$, and Piers Coleman$^2$}
\affiliation{$^1$ Joint Quantum Institute and Department of Physics, University of Maryland, College Park, MD 20742, USA\\
$^2$Center for Materials Theory, Rutgers University, Piscataway, NJ 08854, USA}

\date{\today}

\begin{abstract}
Kondo insulators are particularly simple  type of heavy electron
material, where a filled band of heavy quasiparticles gives 
rise to a narrow band insulator.  Starting with
the Anderson lattice Hamiltonian, we develop a topological
classification of emergent band structures for Kondo
insulators and show that these materials may host three-dimensional
topological insulating phases. We propose a general and practical
prescription of calculating the $Z_2$ strong and weak topological
insulator indices for various lattice structures. Experimental
implications of the topological Kondo insulating behavior are
discussed.
\end{abstract}

\pacs{71.27.+a, 75.20.Hr, 74.50.+r}

\maketitle
%
%
%  Missing: the historic role of KI, SmB6 the first heavy fermion system
%
%  Missing: a clear statement of the idea of adiabaticity - its
%  origins in the work of Allen and Martin.
%
%  Propose that a subset of the KI are infact, topological - adiabatically
%  linked to a non-interacting Kondo insulator
%
%  A new ``resonant'' mechanism for topological insulators. 
%  Resonant Topological Insulators vs Topological Kondo insulators\dots .
%

Kondo insulators are a particularly simple type of heavy fermion
material, first discovered forty years ago\cite{Geballe69},
in which highly renormalized 
f-electrons, hybridized with conduction electrons, form a
completely filled band of quasiparticles with excitation gaps
in the millivolt range \cite{Coleman2007,KIReviews}. 
While these materials are  strongly interacting
electron systems, their excitations and their ground-states can be regarded 
as adiabatically connected to non-interacting band-insulators\cite{allen}.

It was recently shown that time-reversal invariant band insulators
can be classified by the 
topological structure of their ground-state wavefunctions
\cite{Fu2007,Moore2007,Roy2009,Qi2008}. One of the dramatic  consequences of
this discovery, is the existence of a new class of ``topological'' 
band-insulator in which strong spin-orbit coupling leads to a
ground-state that is topologically distinct from the vacuum, 
giving rise to gapless surface excitations. 

In this paper, we show that  Kondo insulators, as adiabatic descendents
of band insulators, can also be topologically classified. 
The strong spin-orbit coupling 
characteristic of these materials leads us to predict that  a subset
of Kondo insulators are topologically non-trivial, with anomalous
surface excitations.  In current models of topological insulators,
the spin-orbit coupling is encoded in a spin-dependent hopping
amplitudes between different unit cells. By contrast, in a 
topological Kondo insulator (TKI), we show that 
the topologically nontrivial insulating state is produced 
by the spin-orbit coupling associated with the 
hybridization between conduction and $f$-electrons. 

Below, we develop a simple
model for topological KIs. The physics we study is
motivated by the canonical Kondo insulating behavior of
SmB$_{6}$\cite{Geballe69} and Ce$_3$Bi$_4$Pt$_3$ \cite{Hundley1990}. 
The realization of a particular topologically
nontrivial insulating state depends on the position of renormalized
$f$-level relative to the bottom of the conduction band (see
Fig. 1). To analyze the topology of the bands in these materials, we
use a periodic Anderson lattice model. 

In a KI, the insulating state arises due to hybridization between the
conduction and $f$-electrons, provided that the chemical potential
lies inside the hybridization gap separating the quasiparticle bands.  The
spatial symmetry of the hybridization amplitude is determined by the
symmetry of the underlying crystal-field Kramers doublets of the rare-earth
ion and it is precisely this symmetry that is responsible for 
non-trivial topological structures in a KI. To analyze
this topology, we first employ a tight-binding model on a simple cubic
lattice, which is adiabatically connected to the Hamiltonian of the KI
material. We show that in this case, band topology is uniquely
determined by the noninteracting band structure of the system in the
absence of hybridization. Secondly, we consider a more general KI in a
lattice with a body centered cubic lattice, and show that regardless of microscopic
details, there always exists a parameter range in which a KI is a
strong topological insulator (STI).

%%%%%% This is Fig. 1 -> TITLE PLOT %%%%%%
\begin{figure}[h]
\includegraphics[width=2.8in]{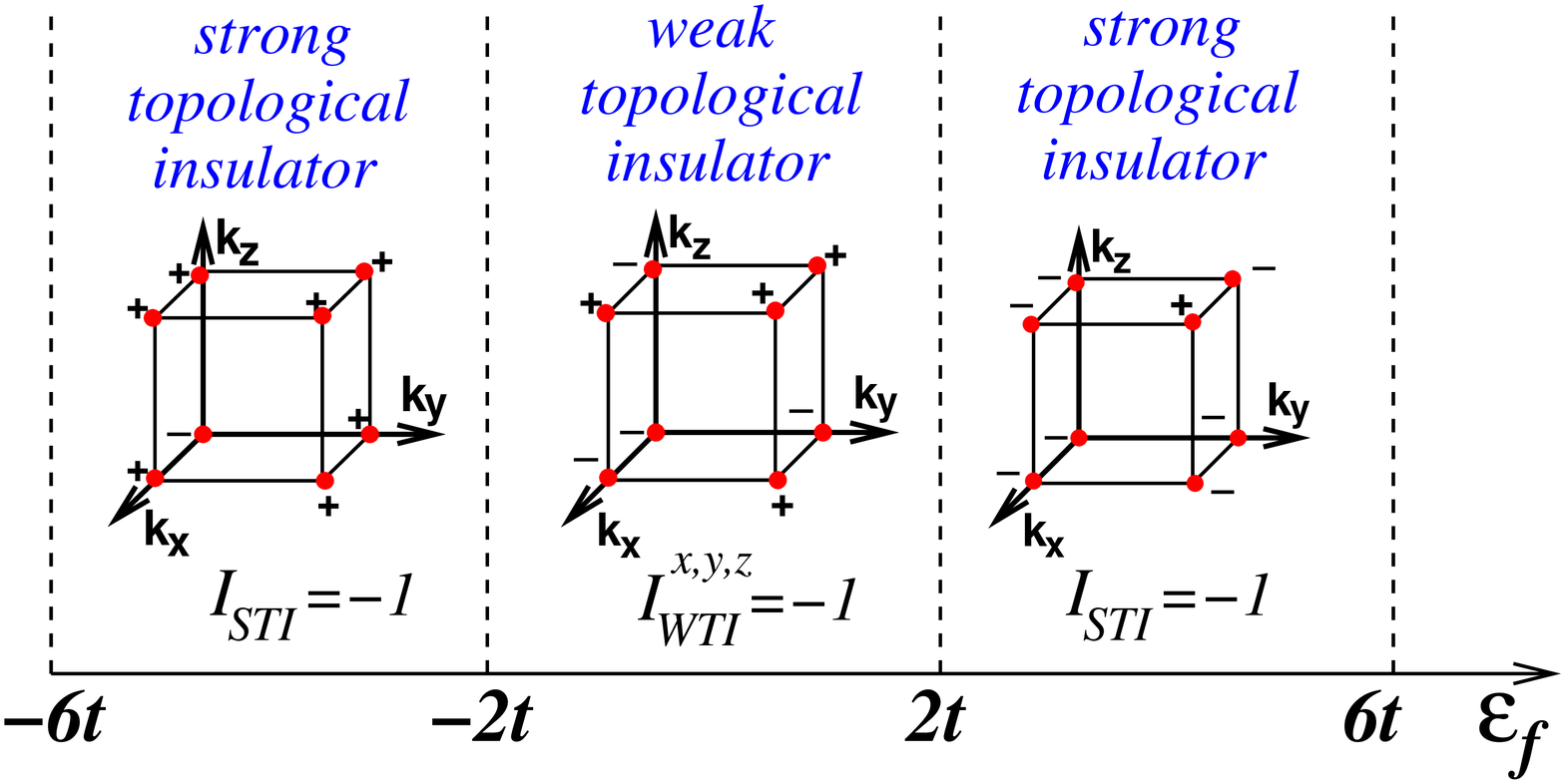}
\caption{Values of strong and weak topological indices and signs of $\delta_i$ (see text) at the high-symmetry 
points of the Brillouin zone (BZ) are shown as a function of position of the renormalized $f$-level 
relative to the bottom of the conduction band. For the simple cubic tight-binding spectrum 
of the form $\epsilon_\bk=-2t\sum_{a=x,y,z}\cos k_a$ topologically strong, weak 
Kondo insulating behavior as well as regular band insulator ($|\varepsilon_f|>6t$) 
can be realized. These results
are shown to be generic and independent of particular type of Bravais lattice. 
%KI is assumed to be due to the $|j=5/2,M=\pm1/2\rangle$-doublet of
%the magnetic ion.
}
\label{Fig1}
\end{figure}
%%%%%% Done with Fig.1 %%%%%%%%%%%%%%%%%%%%%%%

We begin with the periodic Anderson Kondo lattice Hamiltonian, written
in terms of the fermion operators associated with the crystal field
symmetry of the underlying lattice \beg\label{HPALM}
\begin{split}
\hat{H}=&\sum_{\bk , \alpha }\xi_{\bk }c\dg_{\bk \alpha }c_{\bk \alpha }
+
\sum\limits_{j\alpha} \left[V {c}_{j\alpha}\dg {f}_{j\alpha}+ {\rm h.c.}\right]\\ 
&+\sum\limits_{j\alpha}\left[\varepsilon_{f}^{(0)}{n}_{f,j\alpha}+\frac{U_f}{2}{n}_{f,j\alpha}
{n}_{f,j\overline{\alpha}}\right].
\end{split}
\en
where $\xi_{\bk }$ is the dispersion of a tight-binding band of
conduction electrons.
We assume that the ground state of the isolated 
magnetic ion is a Kramers doublet $|\Gamma\alpha\rangle$, where $\Gamma$ labels a particular representation
of the crystal symmetry group. For instance, in a cerium-based Kondo system, the Ce ion has a Ce$^{3+}$ valence configuration, hence we have one $f$-electron in a $J=5/2$ atomic shell. In Eq.~(\ref{HPALM}) above, the operator ${c}_{j\alpha}\dg$ creates an electron on site $j$ in a
Wannier state described by the quantum number $\alpha=\pm$, $\varepsilon_{f}^{(0)}$ is the bare energy of the $f$-level, $V$ is the bare hybridization, and $U_f$ describes Hubbard repulsion between $f$-electrons.
One can relate the Wannier states at site $j$ as follows \cite{Coqblin1969,Flint2008}:
$c _{ j\alpha }=\sum_{k\sigma }[\Phi_{\Gamma\bk}]_{\alpha \sigma }
c_{\bk \sigma }e^{i \bk \cdot{\bf{R}}_{j}}$, where the form factors
$[\Phi_{\Gamma\bk }]_{\alpha \sigma }$ are two dimensional matrices
\begin{equation}\label{eq2}
[\Phi_{\Gamma\bk }]_{\alpha \sigma }=\langle k\Gamma \alpha \vert \bk
\sigma \rangle=
\sum_{m\in [-3,3]}\left\langle \Gamma\alpha \Bigl\vert 3m,
\frac{1}{2}\sigma
\right\rangle \tilde{Y}^{3}_{m-\sigma } ({\bk} ).
\end{equation}
where
\begin{equation}\label{}
\tilde{Y}^{3}_{M} (\bk )= \frac{1}{Z}\sum_{\bf R \neq 0} Y^{3}_{M} (\hat {\bf
R}) e^{i \bk\cdot {\bf R}}
\end{equation}
is a tight-binding generalization of the spherical Harmonics that
preserves the translational symmetry of the hybridization,
$\Phi (\bk )= \Phi (\bk +{\bf G})$,  where $\bf{G}$ is 
reciprocal lattice vector.  Here, 
 $\bf R$ are the positions of the Z nearest neighbor sites
around the magnetic ion.  

The low-energy properties of the model (\ref{HPALM}) are described in terms of
renormalized quasiparticles formed via strong hybridization between the $c-$ and
$f-$ states and on-site repulsion $U_f$. In the regime where the $f$ states are predominantly localized,
we can neglect the momentum dependence of the $f$-electron self-energy
$\Sigma_f(\bk,\omega)\simeq\Sigma_f(\omega)$ so that
the effective low-energy Hamiltonian reads \cite{Ikeda1996}:
\beg\label{Hmf}
\mathcal{H}_{mf}(\bk)=
\left(
\begin{matrix}
\xi_\bk\underline{1} & \tilde{V}{\Phi}_{\Gamma\bk}\dg \\
\tilde{V}{\Phi}_{\Gamma\bk} & \varepsilon_f\underline{1}
\end{matrix}
\right),
\en
where $\xi_\bk$ is the bare  spectrum of conduction electrons  taken relative to the chemical
potential, $\varepsilon_{f}=z[\varepsilon_{f}^{(0)}+\Sigma_f(0)] $  is the renormalized $f$-level, $\tilde{V}=\sqrt{z}V$ , $z=(1-\partial\Sigma_f(\omega)/\partial\omega)_{\omega=0}^{-1}$, and 
$\underline{1}$ denotes the unit $2\times2$ matrix.  
The KI is formed if the chemical potential of the quasiparticles lies inside
the hybridization gap, separating the two bands with the spectra
$E_{\pm}(\bk)=
\frac{1}{2}[\xi_\bk+\varepsilon_f\pm\sqrt{(\xi_\bk-\varepsilon_f)^2+4\left|\tilde{V}\Delta_\bk\right|^2}]$,  with 
$\Delta_\bk^2=\frac{1}{2}\text{Tr}[{\Phi}_{\Gamma\bk}\dg{\Phi}_{\Gamma\bk}]$.

From Eq.~(\ref{eq2}), we see that the form factors $\Phi_{\Gamma \bk
}$ are momentum-dependent unitary matrices that relate the spin
quantization axes of the Bloch states and the spin-orbit coupled
Wannier states.  Our choice of hybridization ensures that 
 the mean-field Hamiltonian (Eq. \ref{Hmf}) is
a periodic function satisfying $\mathcal{H}_{mf}({\bf
k})=\mathcal{H}_{mf}({\bf k}+{\bf G})$.

These form factors are uniquely determined by the
wave-functions of a magnetic ion, $|\Gamma\alpha\rangle$. For example,
in a tetragonal crystal field environment, the $j=5/2$ Ce multiplet is
split into three doublets: $|\Gamma_1^{(t)}\pm\rangle=|\pm1/2\rangle$,
$|\Gamma_2^{(t)}\pm\rangle=\cos(\beta)|\mp3/2\rangle+\sin(\beta)|\pm
5/2\rangle$, and
$|\Gamma_3^{(t)}\pm\rangle=\sin(\beta)|\mp3/2\rangle-\cos(\beta)|\pm
5/2\rangle$, where the mixing angle $\beta$ defines orientation of the
corresponding states. In an orthorhombic environment, the Kramer's
doublets are generally described by a linear superposition of all three
wave-functions $|\Gamma^{(o)}\pm\rangle= u\vert 
|\pm 1/2\rangle+v|\mp 3/2\rangle+w\vert \pm 5/2 \rangle $
\cite{Kikoin1994,Moreno2000}. To discuss the topological properties of
these KI (\ref{Hmf}), we need to consider separately the form factors
for different $\Gamma$'s. We distinguish these states according to
their orbital symmetry parameterized by the index $a=1,2,3$ and the
pseudo-spin quantum number ($\alpha=\pm$) \cite{Moreno2000}. Hence, we
have ${f}_{1\pm}\dg|0\rangle=|\pm1/2\rangle$,
${f}_{2\pm}\dg|0\rangle=|\pm3/2\rangle$, and
${f}_{3\pm}\dg|0\rangle=|\pm5/2\rangle$.

The momentum-dependence of the hybridization gap $\Delta_a(\bk)$
follows from Eq. (\ref{eq2}). At small momenta $\bk $, 
$\Delta_1(\bk)=\frac{1}{12}\sqrt{\frac{3}{\pi}}[12\cos(2\theta)+5(3+\cos(4\theta))]^{1/2}$,
$\Delta_2(\bk)=\frac{1}{8}\sqrt{\frac{3}{\pi}}|\sin\theta|[17+15\cos(2\theta)]^{1/2}$,
and $\Delta_3(\bk)=\frac{1}{4}\sqrt{\frac{15}{2\pi}}\sin^2\theta$,
where $\theta$ and $\phi$ define the direction of the unit vector
$\hat{\bf k}$, associated with the point on the Fermi surface. Note
that the hybridization gap has a line of nodes along the $z$-axis for
the shapes $a=2,3$, but generic combinations of all three
form-factors characteristic of 
contain no nodes. The key results of this paper are most simply illustrated
using the nodeless $a=1$ Kramers doublet as the ground-state
of the magnetic ion. 

To analyze the topology of the bands we  use the fact that 
topology  is invariant under any adiabatic deformation
of the Hamiltonian. 
We begin our study with a 
tight-binding model for a KI on a simple cubic lattice. 
The technical analysis is readily generalized to 
more complicated cases as discussed below.  The most important element 
of the analysis is the odd parity form factor of the 
$f$ electrons,
${\Phi}_{a}(\bk)=-{\Phi}_{a}(-\bk)$.  
This parity
property is the only essential input as far as the topological
structure is concerned.

In Ref.~\cite{FuKane2007}, 
Fu and Kane demonstrate that in an insulator
{\em with time-reversal and space-inversion symmetry}, the topological
structure is determined by parity properties at the eight
high-symmetry points, $\bk^*_m$, in the 3D BZ which are invariant
under time-reversal, up to a reciprocal lattice vector:
$\bk^*_m=-\bk^*_m+{\bf G}$ (see insets in Fig.~\ref{Fig1}).  In our
case, these symmetries require that $\mathcal{H}_{mf}({\bf k})={P}
\mathcal{H}_{mf}(-{\bf k}){P}^{-1}$ and $\mathcal{H}_{mf}({\bf
k})^{T}={\cal T} \mathcal{H}_{mf}(-{\bf k}){\cal T}^{-1}$, where
the parity matrix $P$
and the unitary part of the time-reversal
operator ${\cal T}$ are given by
\begin{equation}\label{}
P = \begin{pmatrix} \underline{1}& \cr & -\underline{1}\end{pmatrix},
\qquad 
{\cal T} = \begin{pmatrix}  i \sigma_{2}&\cr & i \sigma_{2}  \end{pmatrix},
\end{equation}
where $\sigma_{2}$ is the second Pauli matrix. 
For any
space-inversion-odd form factor, it follows immediately that
$\hat{\Phi}_{a}(\bk)=0$ at a high-symmetry point. Hence, the
Hamiltonian at this high symmetry point is simply
$\mathcal{H}_{mf}({\bk^*_m})=(\xi_{\bk^*_m}+\varepsilon_f)
I/2+(\xi_{\bk^*_m}-\varepsilon_f){P}/2$, where $I$ is 
the four-dimensional identity matrix.

The parity at a high symmetry point is thus determined by $\delta_m=\textrm{sgn}(\xi_{\bk^*_m}-\varepsilon_f)$.
Four independent $Z_2$ topological indices~\cite{Kitaev} (one strong and three weak indices) can be
constructed from $\delta_m$: (i)~The strong topological index is the product of all eight $\delta_m$'s: 
$I_{\rm STI} = \prod\limits_{m=1}^{8} \delta_m = \pm 1$; (ii)~by setting $k_j=0$ (where $j= x,y, \mbox{and } z$),  
three high-symmetry planes, $P_j = \left\{ {\bf k}: k_j=0\right\}$, are formed that contain four high-symmetry points each. The product of the parities at these four points defines the
corresponding weak-topological index, $I_{\rm WTI}^j =  \prod\limits_{{\bf k}_m \in P_j} \delta_m = \pm 1$, $j=x,y,z$. The existence of the three weak topological indices in 3D is related to a $Z_2$ topological index for 2D systems (a weak 3D TI is similar to a stack of 2D $Z_2$ topological insulators). 
Because there are three independent ways to stack 2D layers to form a 3D system,
the number of independent weak topological indices is also three.

A conventional band insulator has all of the four indices $I_{\rm STI}
= I_{\rm WTI}^x=I_{\rm WTI}^y=I_{\rm WTI}^z = +1$, while an index
$I=(-1)$ indicates a $Z_2$ topological state with the odd number of
surface Dirac modes. For a KI with $\xi_{{\bk_m^*}=0}<\varepsilon_{f}$
and $\xi_{\bf{{\bk^*_m\ne 0}}}>\varepsilon_{f}$, we find $I_{\rm STI}
= -1$, and hence {\em the Kondo insulating state is a
strong-topological insulator which will be robust with respect to
disorder}. Weak-topological insulators and topologically trivial
insulators can in principle be found for different band structures and
different values of $\varepsilon_{f}$, Fig. 1. A particularly
interesting possibility is to tune topological phase transitions
between different types of insulators (e.g., by applying a pressure),
Fig. 1. Although we have been specifically considering a tight-binding
model with a primitive unit cell, all our conclusions apply directly
to systems adiabatically connected to this model. In order to prove it
explicitly and to investigate more general cases, we develop a
different and more general technique similar to that proposed in
Ref.~\cite{Roy2009}.

We shall now study an example of a KI for the specific shape $a=1$
and explore the parameter range in which it remains a STI.  Here, the form factor is universally
determined in the small-momentum limit by the $f$-wave symmetry of the
electron orbitals and the point group symmetry of the lattice. 
Expression for the form factor of a Kramer's doublet $\Gamma_1$ is
\begin{align}
\hat\Phi_{1\bk}\propto\frac{k^3}{\sqrt{7}}
\left[\begin{matrix}
\sqrt{3}Y^3_0(\hat{\bk}) & -2 Y^3_1(\hat{\bk}) \\
2 Y^3_{-1}(\hat{\bk}) & -\sqrt{3}Y^3_0(\hat{\bk})
\end{matrix}\right],
\label{eq:form_factor_Y}
\end{align}
where $\hat{\bk}=\bk/k$ and $Y^l_m(\hat{\bk})$ are the spherical harmonics.
At larger momenta, the form-factor depends on the microscopic details of the lattice and
the Kondo coupling. In general form factor can be written as
$\hat\Phi_{1\bk}=\vec{m}_\bk\cdot\vec{\sigma}$.
%(while the admixture of 
%$\hat\Phi_{2\bk}$ and $\hat\Phi_{3\bk}$ into the form-factor 
%may introduce an component proportional to the unit vector $i m^{0}\underline{1}$).

\begin{figure}[h]
\includegraphics[width=2 in]{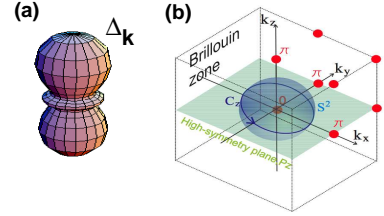}
\caption{(a) Surface plot of the hybridization gap $\Delta_{\bk}$ for the form factor of shape $a=1$, Eq. (3).
The $k$-dependence of the hybridization gap originates from the underlying orbital structure of the
localized electrons, which form the magnetic moment.
(b) The BZ of an orthorhombic lattice (or lattice with a higher point-group symmetry). The dots at the origin, face centers, bond centers,
and the corner  mark the eight high-symmetry points. The sphere at the origin separate regions I (inside) and II
(outside). The intersection of the sphere with the high symmetry plane ($k_z=0$) is marked by the solid line.}
\label{Fig2}
\end{figure}

We show below that momenta where $\hat\Phi_{1\bk}=0$ are crucial
in calculating the topological indices, so that our results are generic
for a linear combination of all three shapes. The most obvious zero-point is located
in the origin as shown in Eq.~\eqref{eq:form_factor_Y}. We now prove that this zero-point
is topologically protected and that its existence necessarily yields the existence
of other zero-points $\hat\Phi_{1\bk_i}=0$ with $\bk_i\ne 0$  (this conclusion is similar to fermion
doubling of relativistic fermions, which requires the presence of an even number of Dirac points on a compact manifold). To prove this, we draw a sphere $S^2$ at the origin with radius $k_0$ as shown in Fig.~\ref{Fig2}(b)
and require $\hat\Phi_{1\bk}\ne 0$ on the sphere, which enables the definition of
a Chern number
$C=\frac{1}{8\pi}\oint_{S^2} \vec{dS} \cdot \epsilon^{ijk}n_i (\vec{\nabla}_{\bk} n_j)\times
( \vec{\nabla}_{\bk}n_k)$, 
where $n_i=m_i/\sqrt{m_x^2+m_y^2+m_z^2}$ and $m_i$ is the $\hat{\sigma}_i$ component
of $\hat{\Psi}_1$ as defined above. The Chern number is a topological index quantized to an integer value.
For small $k_0$, Eq.~\eqref{eq:form_factor_Y} is asymptotically accurate, which gives
$C=1$. Notice that this topological index is invariant as $k_0$ changes
adiabatically. The non-zero value of $C$ indicates that $k_0$ cannot be decreased to zero smoothly,
and hence a point with $\hat\Phi_{1\bk}=0$ must exist inside the sphere, which ensures
$\hat{\Phi}_{1\bk=0}=0$. Since the BZ has a periodic structure and is a compact manifold,
the same argument requires zero points $\hat\Phi_{1\bk_m}=0$ outside the sphere with
$|\bk|_m>k_0$ ($m=1,2...$). This conclusion can be verified explicitly in the simple cubic lattice model
studied above, where $\hat{\Phi}_{1\bk}=0$ at each of the eight high symmetry points.

Now we relax the assumption about the simple cubic lattice and allow for a more general structure. 
Due to time-reversal and space-inversion symmetries, both bands are doubly-generated in a KI.
Therefore, the corresponding Bloch wavefunctions $\Psi^{1}(\mathbf{k})$ and
$\Psi^{2}(\mathbf{k})$ (which are four-component vectors)
can be chosen arbitrarily up to a ``local'' $U(2)$ transformation in momentum space.
For fermions, ${\cal T}^2=-I$, and hence, we require that 
${\cal T} \Psi^{i}(\mathbf{k}) = \epsilon_{ij}[\Psi^{j}(-\mathbf{k})]^*$ under time reversal
with $\epsilon_{ij}$ being the Levi-Civita symbol.
For concreteness, we focus below on the case with $\xi_\bk<\varepsilon_f$ at $\bk=0$ and
$\xi_{\bk_i}>\varepsilon_f$ at all other zero points of $\hat{\Phi}_{1\bk}$, which gave us
a strong topological insulator in the model with a simple cubic lattice discussed above.
For such a band structure, it can be easily checked that the wave-functions can {\em not} be defined
globally in the entire BZ with the constraint
${\cal T} \Psi^{i}(\mathbf{k}) = \epsilon_{ij}[\Psi^{j}(-\mathbf{k})]^*$ \cite{Fu2006}.
However, the BZ can be separated into two regions (c.f., Ref.~\cite{KSVG}) $I$ ($|\bk|<k_0$) and $II$
($|\bk|>k_0$) with $k_0$ being an arbitrarily small momentum such that only one zero point of $\hat{\Phi}_{1\bk}$, $\bk=0$ is enclosed inside the sphere, Fig.~\ref{Fig2}(b).
In each of the two regions, singularity-free Bloch wave-functions can be constructed. For example,
the valence band has following wave-functions in region $(I)$
\begin{equation}\label{Psi1}
\begin{split}
&\Psi^{1}_I(\mathbf{k})=\mathcal{N}_-(0,E_-,-\Phi_{12},-\Phi_{22}), \\
&\Psi^{2}_I(\mathbf{k})=\mathcal{N}_-(E_-,0, -\Phi_{11}, -\Phi_{21}),
\end{split}
\end{equation}
and in region $(II)$
\begin{equation}\label{Psi2}
\begin{split}
&\Psi^{1}_{II}(\mathbf{k})=\mathcal{N}_+(-\Phi_{21}^*,-\Phi_{22}^*,0,E_+), \\
&\Psi^{2}_{II}(\mathbf{k})=\mathcal{N}_+(-\Phi_{11}^*,-\Phi_{12}^*,E_+,0).
\end{split}
\end{equation}
Here $\Phi_{ij}$ are the $(i,j)$-components of the form-factor,
$E_{\pm}(\bk)$ and  $\Delta_\bk$ were defined above Eq.~(\ref{Hmf}), and
$\mathcal{N}_{\pm}=1/\sqrt{2\tilde{V}^2\Delta_\bk^2\pm(\xi_{\mathbf{k}}-\varepsilon_{f})E_{\pm}}$.
These two sets of Bloch wave-functions are connected by a ``gauge'' transformation
$\Psi^{i}_I(\mathbf{k})=\hat{U}_{ij}(\mathbf{k}) \Psi^{j}_{II}(\mathbf{k})$
at the boundary, $S^2$, between the regions $I$ and $II$, with the matrix
$\hat{U}(\bk)=-\hat\sigma_z\hat{\Phi}_{1-\mathbf{k}}^\dagger\hat\sigma_z/\gamma$.

The topological structure of a 3D time-reversal invariant insulator is determined by the wavefunctions
(\ref{Psi1},\ref{Psi2}) on the six high-symmetry planes 
$P_j = \left\{{\bf k}: k_j=0\right\}$ and $P'_j = \left\{{\bf k}: k_j=\pi\right\}$
~\cite{FuKane2007,Moore2007,Roy2009}. As shown in Fig.~\ref{Fig2}(b), the boundary 
between the two regions intersects with $P_j$ and the intersections are circles, $\mathcal{C}_j$.
On such a circle, the matrix $\hat{U}$ takes the form of 
$\hat{U}(\bk)=exp(i\varphi_0 \hat{\sigma}_0+i \varphi_{\bk}\hat{\bm n}\cdot\hat{\bm \sigma})$, 
with $\hat{n}$ being a fixed 3D unit vector, corresponding to a specific ``gauge.''
As a result, a winding number can be defined on $\mathcal{C}$ as
$w_j= \oint_\mathcal{C_j} \frac{d\mathbf{k}}{2\pi} \cdot\nabla_\mathbf{k} \varphi_\bk$.
However, it is defined modulo $2$ only, because a gauge transformation can change
$\varphi_\bk\rightarrow\varphi_\bk+2 m \phi$ and hence change $w_j$ by an even number $2 m$,
where $\phi$ is the azimuth angle of $\mathbf{k}$ in a high symmetry plane (note that a transformation $\varphi_\bk\rightarrow\varphi_\bk+(2 m+1) \phi$ 
would violate the symmetry constraint ${\cal T} \Psi^{i}(\mathbf{k}) = \epsilon_{ij}[\Psi^{j}(-\mathbf{k})]^*$).
For $P'_j$, the corresponding winding numbers $w'_j$ are zero for the case we studied here,
because they do not intersect with the boundary.
The  topological indices can be computed from these winding numbers as follows~\cite{multi}:
\begin{equation}
\label{Tind}
I_{\rm STI} = \left(-1\right)^{w_j+w'_j} \mbox{ and } I_{\rm WTI}^{j} =  \left(-1\right)^{w_j},
\end{equation}
where $I_{\rm STI}$ is equivalently defined for $j=x$, $y$ or $z$.
For the types of KI band structures considered here, the topological indices can be universally determined by choosing a small enough $k_0$ and using the asymptotic form-factor of Eq.~\eqref{eq:form_factor_Y}.
As a result, we find that the generic Kondo system is a STI in full agreement with arguments above based on adiabatic deformation of the Hamiltonian onto a simple cubic lattice. 

Let us briefly discuss the implications of our results for existing Kondo insulators. 
From our theory's we expect that materials in which $f$-electrons
are close to integral valence are likely to be weak topological Kondo insulators and thus
are unstable with respect to disorder. An interesting example is SmB$_6$ for which 
recent LSDA+$U$ band structure calculations \cite{Harmon2002} show
the position of the $f$-level equals approximately one sixth of the bandwidth 
consistent with the core-level spectroscopy measurements of the $f$-level occupation $n_f\sim 0.7$
\cite{Chazalviel1976}, placing the quasiparticle f-level of SmB$_6$ close
to the border separating STI and WTI phases. 
%A compelling possibility would be to see whether an application of 
%external or chemical pressure can lower the position of the $f$-level and 
%induce the transition from WTI to STI phase in this material. 
Another promising candidate for the manifestation of topologically
nontrivial insulating state is CeNiSn.  Recent transport data in
CeNiSn \cite{CeNiSn2002} shows suppression of semiconducting behavior
in resistivity with increase in sample's quality, although there is an
evidence for the gap formation at $T\simeq 10$K. Given that the
$f$-electrons in these systems are predominantly localized, it is
tempting to speculate that the CeNiSn is a weak topological Kondo
insulator ascribing the semi-metallic transport properties 
to metallic surface states. These are issues that 
we hope can be resolved in the near future through more accurate
modelling and the use of high precision ARPES and STEM spectroscopy. 

To summarize, we have developed a theory of topological 3D
Kondo insulators. Within our model topologically nontrivial insulating states
are realized over a wide parameter range. In particular, we have shown 
that strong topological insulating state occurs when the
position of the re-normalized $f-$ level is near the top, or the bottom of the
conduction band. This suggests the most likely candidates for this kind of behavior 
are heavy fermion materials which are more mixed valent or have narrow conduction bands. 

This work was supported by DARPA (M. D. and V. G.),  JQI-NSF-PFC
(K. S.), NSF-CAREER (V. G.), and DOE grant DE-FG02-99ER45790 (P. C.).
We would like to thank Joel Moore and David Vanderbilt 
for discussions related to this project.

%\bibliography{TKI}

\end{document}